\documentstyle[aps,preprint]{revtex}
\begin{document} 
\draft 
\preprint{{IC/97/161;TIFR/TH/97-59;IMSc-97/11/40;}\\
{hep-th/9801194}} 
\title{Planckian Scattering of D-Branes} 
\author{Saurya Das$^{1,2}$ \footnote{E-mail: saurya@imsc.ernet.in},
Arundhati Dasgupta$^1$ \footnote{E-mail: dasgupta@imsc.ernet.in}, 
P.Ramadevi$^3$ 
\footnote{E-mail: rama@theory.tifr.res.in}
and Tapobrata Sarkar$^1$ \footnote{E-mail: sarkar@imsc.ernet.in}}
\address{$^1$~The Institute of Mathematical Sciences, CIT Campus, 
Chennai - 600113, India \\
$^2$ International Centre for Theoretical Physics, P.O. Box 586, 34100
Trieste, Italy \\
$^3$~Tata Institute of Fundamental Research, Homi Bhabha Road, Mumbai - 400 005, India
} 
\maketitle 
\begin{abstract}
We consider the gravitational scattering of point particles in
four dimensions,
at Planckian centre of mass energy and low momentum
transfer, or the eikonal approximation. 
The scattering amplitude can be exactly computed by
modelling point particles by  
very generic metrics. 
A class of such metrics are black hole 
solutions obtained from dimensional reduction of p-brane
solutions
with one or more Ramond-Ramond charges in string theory. 
At weak string coupling, such black holes are replaced by
a collection of wrapped $D$-branes. 
Thus, we investigate eikonal scattering at weak coupling by modelling
the point particles 
by wrapped $D$-branes and show that the amplitudes 
exactly match the 
corresponding amplitude found at strong coupling. 
We extend the calculation for scattering of charged particles. 

\end{abstract}

\section{Introduction}

It is expected that the quantum gravity effects would become important at
energies compared to the Planck Scale. 
Since the gravitational coupling constant $G$ is not dimensionless,
one can construct two {\it independent dimensionless} coupling constants, 
which for example in four space-time dimensions can
be defined as $G_{\parallel} \equiv G_4 s$ 
and $G_{\perp} \equiv G_4 t$. Here $G_4$ is the four
dimensional Newton's constant, and $s,t$ are the Mandelstam variables.
Remarkably, it was shown in \cite{vv} that the full theory of quantum
gravity can be split up into two independent theories with these coupling
constants. Thus, a quantum gravitational regime
for gravitational scattering can be envisaged
when either $G_{\parallel} \approx 1$ or $G_{\perp} \approx 1$, or both.
While the former
signifies Planckian centre-of-mass energies, the latter implies Planckian
momentum transfers. The first quantum gravitational
scenario is easier to deal with because
although $s$ is large, $t$ can be held fixed at a relatively small value
such that $s/t \rightarrow \infty$.
The impact parameter of scattering, in this case, is very large and the
scattering is almost forward.
This is the so called eikonal approximation where the {\em exact}
two particle scattering amplitude can be computed. 
In practice, it is advantageous to view one of the particles as
static (say $A$) and the other moving (say $B$) 
at almost luminal velocity past it with a
large impact parameter. $A$ being static, can be suitably modelled by a
metric, whose gravitational field $B$ is supposed
to experience. Then one can solve the wave equation for $B$ 
in this given background and obtain the scattering
amplitude. Of course, the reverse process is equally valid, when
the
`shock-wave' space time produced by the $A$ is obtained by
Lorentz boosting
the metric and analysing the wave function of the slow particle
in this shock-wave background. As expected, the two
pictures yield the same result \cite{th}. 

In the above picture, the point particles are usually modelled by
Schwarzschild or Reissner-Nordstr\"om metrics, depending on whether the
particles are neutral or charged. Although this seems natural in the
framework of general relativity, these specific choices are certainly
not mandatory. We show here that the results can be extended
for a large class of generic spherically symmetric metrics. As
an example, a
large class of metrics arise as solutions of low energy string theory
and one could model the particles by these metrics as well.
The black holes which carry the NS-NS charges have already been considered in Planckian scattering which
lead to interesting consequences \cite{DM}. It is shown here that
these black hole metrics do not fall into
the class of metrics we consider here, except in the extremal
limit. Recently a new class
of black hole solutions of low energy effective action of
 superstring theory have
been found, whose weak coupling (in terms of the string coupling $g$)
description consists of certain configurations of solitonic string states
or D-branes, wrapped on suitable compact manifolds. Several pieces of
evidence have emerged supporting this
identification, the most notable being the fact that the degeneracy associated with
the D-brane
configuration exactly reproduces the Bekenstein-Hawking entropy \cite{VS}
and open string interactions on the $D$-brane reproduce the 
Hawking radiation spectrum of these black holes \cite{haw}. Thus, while the
black hole description can be used at large coupling, 
the $D$-brane
description is appropriate at small coupling 
\cite{VS,haw,horpol}. 
In \cite{bac,L1,L2,BL}, it has been shown scattering amplitudes of R-R charged 
p-branes agree with appropriate D-brane scattering in ten
dimensions.
There are black hole solutions with a singular horizon
obtained by wrapping these R-R charged p-branes on compact spaces.
The entropy for these black holes is zero, and hence the
appropriate process to check for D-brane black hole
correspondence is to look at scattering amplitudes. 
In this paper, we show that indeed the exact eikonal scattering amplitude can
be computed for wrapped D-branes at weak coupling. Moreover, this
amplitude agrees with that found in the black hole picture. The
agreement persists when the particles carry $U(1)$ charges also.

In the following section, we calculate the eikonal scattering phase
shift in the strong coupling regime by modelling the particles by  a
general spherically symmetric black hole metric in four space-time
dimensions. We also consider the special cases of R-R and
NS-NS charged black holes. In the next  
section, we calculate the corresponding phase shift at weak coupling using 
$D$-$p$-branes wrapped on tori. 
The $D$-brane result is found to be 
independent of the brane dimensions as long as they are completely
wrapped on the internal tori.
Significantly, the scattering phase shift is dominated by graviton
exchanges at ultrarelativistic velocities, 
as anticipated earlier in the calculations of
\cite{th}. Finally, we extend the calculations to include particles
carrying electric charges, where too the results continue to agree.

\section{Eikonal Scattering at large string Coupling}
\label{large}

We will assume that the slow target particle gives rise to the following
most general spherically symmetric metric in four dimensions :
\begin{equation}
ds^2~=~-\lambda^2dt^2 + \frac{1}{\lambda^2} dr^2 + R r^2 d\Omega^2~,
\label{mas}
\end{equation}
where $\lambda$ and $R$ are functions of the radial coordinate only.

In general let the ultrarelativistic particle of charge $e$ 
be minimally coupled
to the $U(1)$ gauge field $A_\mu$ produced by the static particle.
Then its wave function $\Phi$ 
satisfies the covariant Klein-Gordon Equation
\begin{equation}
\frac{1}{\sqrt -g}~D_\mu~\left( {\sqrt {-g}} g^{\mu
\nu}~D_\nu \Phi
\right)~+~m^2\Phi ~=~0 ~;~~D_{\mu}= \partial_{\mu}- ie A_\mu 
~,
\end{equation}
where $m$ and $E$ are the mass and energy of the test particle.
In the spirit of \cite{th} we will assume that $M,m \ll M_{pl}$, where
$M_{pl}$ is the Planck mass. In the centre of mass frame of
the particles, their energies are proportional to $\sqrt s$
and for the particles to
have relativistic velocities at this energy $M,m\ll \sqrt s$.
Since in our problem $\sqrt s \sim M_{pl}$ we get the above condition on the mass. 
Thus we will ignore all terms quadratic in
$M,m$ in subsequent discussions.
We substitute a solution of the form
$$ \Phi (\vec r, t) ~=~ \frac{\Phi'}{r}~e^{iEt}
Y_{lm}(\theta,\phi)~,$$
and linearize the metric components at large distances as:
$$\lambda^2~=~1 - \frac{2GM}{r}~~,$$
$$R~=~1+ {\cal{O}}(1/r)~.$$
Here $M$ is the ADM mass of the black hole metric. Retaining terms
upto order $1/r^2$, the radial equation of the $\bar{l}^{th}$ partial wave
is (for large $\bar l$) :
\begin{equation}
\frac{d^2\Phi'}{dr^2}  - \left[ \frac{\bar l(\bar l+1)}{r^2} - E^2 -
\frac{4G_4ME^2}{r} + 2eE A_0 \right] \Phi'~=~0~.
\end{equation}
The gauge potential is assumed to have $A_0$ as the only non-zero
component. With its explicit spherically symmetric form 
\begin{equation}
A_0= K/r 
\label{amu}
\end{equation}
and 
the identity $s=2ME$, the above equation reduces to:
\begin{equation}
\frac{d^2\Phi'}{dr^2}  - \left[ \frac{\bar l(\bar l+1)}{r^2} - E^2 -
\frac{2(G_4s-eK)E}{r} \right] \Phi'~=~0~.
\end{equation}

It is straightforward to obtain the phase shift
from here and the answer is \cite{lan}:
\begin{equation}
\delta_{\bar l}~=~\arg \Gamma({\bar l} + 1 - iG_4 s)~.
\end{equation}
Expanding the rhs for $ \bar l \gg 1$, we get \cite{abra}:
\begin{equation}
\delta_{\bar l}~=~-(G_4s -eK)~\ln \bar l~.
\label{gs}
\end{equation}

The above phase shift resembles Rutherford scattering 
with the fine structure constant $\alpha$ being replaced by the
effective coupling constant $-(G_4s -eK)$, which is attractive for
large $s$. The phase shift can be substituted in 
\begin{equation}
f(s,t)~=~\frac{1}{2i {\sqrt s}} \sum_{\bar l=0}^\infty (2\bar l+1) \left[ e^{2i
\delta_{\bar l}}-1 \right]~P_{\bar l} (\cos \theta)~
\end{equation}
to obtain the scattering amplitude. Using the asymptotic formula for
large $\bar l$ 
$$ P_{\bar l} (\cos \theta) \rightarrow J_0 \left( (2\bar l+1) \sin \theta/2
\right)~~,$$
and converting the sum into an integral as in \cite{vv}, we get:
\begin{equation}
f(s,t)~=~-i {\sqrt {s}} \int_0^{\infty} dy~y^{1-2iGs}~J_0\left(2y{\sqrt{s}}
\sin \frac{\theta}{2}\right)~~,
\end{equation}
where $y \equiv\bar  l/{\sqrt{s}}$ and $t= 4E^2\sin^2\theta/2$. Finally, one gets:
\begin{equation}
f(s,t)~=~\frac{-i(G_4S -eK)}{\pi t}~\frac{\Gamma(1-i(G_4s - eK))}
{\Gamma(1+i(G_4 s -eK))}~
\left(\frac{4}{-t}\right)^{-i(G_4s - eK)}~~.
\label{amp}
\end{equation}
The cross section follows:
\begin{equation}
\sigma (s,t)~=~\frac{4(G_4s-eK)}{t^2}~.
\label{cs}
\end{equation}
The amplitude (\ref{amp}) exhibits the infinite set of 't Hooft poles at
the values $G_4s - eK = -iN$, where $N=1,2,...,\infty$. 
Note that for ordinary particles, $eK \ll 1$ and the electromagnetic
contribution to the scattering is suppressed at Planckian energies.
In other words, gravity assumes the role of the dominant interaction
at the Planck scale. 

The linearizations of the functions $\lambda$ and $R$ are valid for 
spherically symmetric metrics of general relativity as well as black
holes carrying Ramond-Ramond charges. Examples in four
dimensions are \cite{gk}: 
\begin{equation}
ds^2~=~-f^{-1/2} h dt^2 + f^{1/2} \left( h^{-1} dr^2 + r^2 d\Omega^2
\right) ~~,
\label{met}
\end{equation}
where $h= 1 - r_0/r$, $f = \prod_{i=1}^4 \left( 1 + r_i/r\right)$ and 
the horizon is at $r_0$. This black hole metric arises when
three distinct five branes of $M$-theory intersect along a line which is
then wrapped on a circle. The parameters $r_i~, i= 1,..,4$ are related to
the four $U(1)$ charges carried by the black hole, three of which are
proportional to the number of the three different $5$-branes, while the
fourth is proportional to the Kaluza-Klein momentum along the intersection
line. The ADM mass of this black hole is $M=(\sum_{i=1}^4~{r_i} +
2r_0)/4G_4$.
For solutions obtained by wrapping BPS saturating fundamental
strings or R-R charged p-branes, the general solution is of
the form {\cite{malth}}:
\begin{equation}
ds^2~=~-f_{p}^{-1/2}  dt^2 + f_p^{1/2} \left( dr^2 + r^2 d\Omega^2
\right) ~~,
\end{equation}
where $f_p= 1 + c_p/r$, $c_p$ is a constant related to the
mass of the p-brane. The horizon is at $r=0$, and thus is
singular in nature. But this metric is well behaved at 
large distances and gives the exact scattering amplitude as
shown in the above calculation.

On the other hand, certain black holes carrying NS-NS charges have
non-extremal metric of the form \cite{GHS}
\begin{equation}
ds^2~=~-\left(1 - \frac{2GM}{r}\right) dt^2 + \frac{dr^2}{\left( 1 -
\frac{2GM}{r} \right)} + r^2 \left( 1 - \frac{\alpha}{Mr} \right)
d\Omega^2~,
\end{equation}
where $\alpha \equiv Q^2 e^{-2\phi_0}$, $Q$ being the electric charge
and $\phi_0$ the asymptotic value of the dilaton field. There is a
curvature singularity at the horizon $r=\alpha/M$, which expands
without limit for vanishing masses. It can be seen that the
corresponding $R(r)$ (Eqn. \ref{mas}) here cannot be linearised in the
asymptotic ($r\rightarrow\infty$) region, except in the extremal limit. 
Thus, the eikonal scattering can be computed with these
metrics only in the extremal limit \cite{DM}.

\section{Eikonal Scattering at small string coupling}

In this section, we will compute the eikonal phase shift at weak string
coupling, when the $D$-brane picture is appropriate. We develop a
general formalism for the scattering of wrapped $D$-branes before
specialising to four dimensions.

Consider a $D$-$p$-brane moving with a relative velocity $v$ with
respect to a $D$-$l$-brane in 10 space-time dimensions. They are
separated by a large
transverse distance $b$. We assume $l \le p$ and that none of the
coordinate directions of the $D$-$l$-brane are orthogonal to those of the
$D$-$p$-brane. Apart from the direction of velocity and the time
coordinate, the end points of an open string ending on
the two branes satisfy either Neumann (N) or Dirichlet (D) boundary
conditions. We denote as $NN$ the number of string coordinates which satisfy
$N$ condition at both the ends. Similarly $ND$ and $DD$. Evidently,
$DD=8-p$, $NN=l$ and $ND= p-l$  
The scattering phase shift
between these two branes are given by the one loop vacuum superstring
amplitude \cite{bac,L1}:
\begin{equation}
\delta~(b)=~\frac{1}{2}\int_0^{\infty}~\frac{d^{NN}k}{(2\pi)^(NN)} \sum_i
\int_0^\infty~\frac{dt}{t}~e^{-2\pi \alpha' t(k^2 +M_i^2)}~~,
\end{equation}
where 
$$M_i^2~=~\frac{b^2}{4\pi^2\alpha'} + \frac{1}{\alpha'} \sum
(\mbox{oscillators})~.$$
The oscillator sum and the integral finally yields
\begin{equation} 
\delta (b)~=~\frac{1}{4\pi}~\int_0^\infty~\frac{dt}{t}~(8 \pi^2 \alpha'
t)^{-NN/2}~e^{-b^2t/2\pi \alpha'} \cdot \left(B \times J \right)~~, 
\label{t}
\end{equation} 
where $B,~J$ are the bosonic and fermionic
contributions to the oscillator sum in the in the one loop open supersting
amplitude given by \cite{L1}
\begin{eqnarray}
B~&=&~f_1^{-(NN+DD)}(q)
f_4^{-ND}(q)~\frac{\Theta_1'(0,it)}{\Theta_1(\epsilon t,it)}~~, \\
J~&=&~\frac12 \left[-f_2^{NN + DD}(q)f_3^{ND}(q) \frac{\Theta_2(\epsilon t, it)}{\Theta_2(0,it)}
+ f_3^{NN+DD}(q)f_2^{ND}(q) \frac{\Theta_3 (\epsilon t, it)}{\Theta_3 (0,it)}\right.\nonumber \\ 
&~&~~~~~~~~~~~- \left. f_4^8(q) \frac{\Theta_4(\epsilon t, it)}{\Theta_4(0,it)} 
\delta_{p,l} \right]~~, 
\end{eqnarray}
where $q \equiv e^{-\pi t}$,
\begin{eqnarray}
f_1(q)~&=&~q^{1/12} \prod_{n=0}^{\infty} (1 - q^n)  \\
f_2(q)~&=&~\sqrt{2} q^{1/12} \prod_{n=0}^\infty ( 1 + q^n ) \\
f_3(q)~&=&~q^{-1/24} \prod_{n=0}^{\infty} ( 1 + q^{2n-1}) \\
f_4(q)~&=&~q^{-1/24} \prod_{n=0}^{\infty} ( 1 - q^{2n-1})~~. 
\end{eqnarray}
The rapidity $\epsilon$ is defined
as $\tanh \pi \epsilon = v$. Note that the last term in $J$
comes from summation of the $NS(-1)^F$ sector and contributes
only when $ND=0$. This term is due to R-R exchange, and we
will concentrate on this term in the next section when we look
at charged particle amplitudes.

Now, to compare the $D$-brane results with the results on the black hole
side, we have to
compactify the branes on suitable compact manifolds, such that in the
non-compact space-time they look like point particles. For
simplicity, we compactify on a $c$ dimensional torus (with $p \le c$),
such that the resultant
noncompact space time is a $(10-c)$ dimensional with a Lorentzian
signature. We take the range of each coordinate of the torus to be $L_i, i=1,...,c$.
Thus, the volume of the torus is $V=\prod L_i$. 
To obtain the phase shift for these wrapped $D$-branes, 
the formula (\ref{t}) has to be modified. For each
compactified $NN$ direction, the momentum integral is replaced
by a
discrete sum in the one loop amplitude. In other words  the
momentum integral is restricted to the remaining non-compact $NN$
directions, and  
and a factor of
$\Theta_3(0,8\pi^2\alpha' t/L^2)$ is inserted in the integrand
 for each compactified direction. Similarly, for each
compact $DD$ direction, a sum over winding modes is introduced, resulting
in a factor of $\Theta_3(0,itL^2/2\pi^2 \alpha')$ in the
integrand \cite{L2}. Since all the $NN=l$ coordinates are compactified and
there are $c-p$ compact $DD$ coordinates, the final result is
\begin{equation} 
\delta~(b)=~\frac{1}{4\pi}~\int_0^\infty \frac{dt}{t}
e^{-b^2 t / 2\pi\alpha'}~
[\prod_{i=1}^l\Theta_3(0,8i\pi^2\alpha' t /L_i^2)]~[\prod_{i=p+1}^c\Theta_3 (0,it L_i^2
/2\pi \alpha')] 
\cdot (B \times J) ~~.
\end{equation} 
Now, large impact parameter $(b \rightarrow \infty)$, scattering is
dominated by the exchange of massless closed string states, for which it
is sufficient to restrict the integrand in the regime $t \rightarrow 0$
\cite{BL,DKPS}. 
Using the relevant formulae given in \cite{L2,BM}, we get:
\begin{eqnarray}
&\Theta_3& (0,8i\pi^2\alpha' t/L) \rightarrow
\frac{L}{2^{3/2}\pi{\sqrt{\alpha'}}}\frac{1}{{\sqrt{t}}}~~, \\
&\Theta_3&(0,itL^2/2\pi^2\alpha') \rightarrow
\frac{2^{1/2}\pi{\sqrt{\alpha'}}}{L}~\frac{1}{\sqrt{t}}~~,\\
&B& \rightarrow
2^{-(p-l)/2}~\pi~t^{[3-(p-l)/2]}~e^{2\pi/3t}~\frac{e^{-\pi
\epsilon^2 t}}{\sinh \pi \epsilon}~~,  \\ 
&J& \rightarrow 4~e^{-2\pi/3t}~e^{\pi
\epsilon^2 t}~\left[2-(p-l)/2 + \sinh^2 \pi \epsilon -2 \delta_{p,l}
\cosh \pi\epsilon \right]~~, 
\end{eqnarray} 
and the phase shift becomes,
\begin{equation} 
\delta(b) ~=~~ \Lambda~\kappa (\epsilon)~
\int_0^\infty \frac{dt}{t}e^{-b^2 t/2\pi \alpha'}
\frac{t^3}{t^{c/2}}~~, 
\label{ps} 
\end{equation} 
where.
\begin{eqnarray}
\Lambda~&=&~\frac{2^{c/2}~\pi^{c-(p+l)}
{\sqrt{\alpha'}}^{c-(p+l)}}{2^{p+l}}\left(\frac{\prod_{i=1}^lL_i}
{\prod_{j=p+1}^cL_j}\right) ~, \label{lambda}
\\
\kappa (\epsilon)~&=&~\frac{2 - (p-l)/2 + \sinh^2 \pi
\epsilon - 2\delta_{p,l} \cosh \pi\epsilon}{\sinh \pi \epsilon}~~.
\label{kappa}
\end{eqnarray}
Note that the integral in the above expression is
independent of $p$ and $l$ and depends only on $c$. That is, it is the
same for branes of arbitrary dimensions for a given compactification.
Thus, scattering phase shifts for all $p,l$ can be calculated from the
above expression provided the branes completely wrap on the internal
torus. Also, as expected, $K(\epsilon=0)=0$ for $p=l$ or $p-l=4$. This is
the familiar no-force condition for BPS states. 

However, for our present purposes, we specialise to the case of $c=6$,
i.e. scattering in 4 dimensional non-compact space time. 
Then the integral over $t$ in (\ref{ps}) simply yields a factor
$-2\ln (b/{\sqrt {2\pi \alpha'}})$.  In addition, we make the
ultrarelativistic approximation $ v \rightarrow 1~,~\epsilon \rightarrow
\infty$, such that 
$$\kappa (\epsilon) \rightarrow \frac12 e^{\pi \epsilon}~.$$
Then Eq.(\ref{ps}) becomes:
\begin{equation}
\delta(b)~=~-\frac{\prod_{i=1}^l L_i ~\prod_{j=1}^p L_j ~e^{\pi\epsilon}}{2^{p+l-3}\pi^{p+l-6}{\sqrt{\alpha'}^
{p+l-6}}~V}~\ln \frac{b}{\sqrt{2\pi\alpha'}}~.
\label{ps1}
\end{equation}
Rewriting this amplitude in terms of the masses of the branes given
by:
\begin{equation}
m_{p(l)}~=~\frac{\prod_{i=1}^{p(l)}L_i}{g (2\pi)^{p(l)} {\sqrt{\alpha'}}^{p(l)+1}}~, 
\label{mass}
\end{equation}
and the Newton's constant in $(10-c)$ dimensions (for $c=6$)
\begin{equation}
G_{10-c}~=~\frac{8\pi^6 g^2 \alpha'^4}{V}~~
\label{G}
\end{equation}
yields
\begin{equation}
\delta(b)~=~-G_4 m_p m_l e^{\pi\epsilon} \ln
\frac{b}{\sqrt{2\pi\alpha'}}~~.
\end{equation}
Finally, using $s=m_p m_l \exp(\pi\epsilon)$, which simply
expresses the relativistic transformation of energy, and $\bar l=bE$, the
amplitude takes the form 
\begin{equation}
\delta~(b)=~-G_4s \ln \frac{\bar l}{E\sqrt{2\pi\alpha'}}~~,
\end{equation}
(the factor $E\sqrt{2\pi \alpha'}$ is irrelevant as it does not
appear in the scattering amplitude and the cross section, and will
simply be dropped).

Thus not only is the phase shift exactly calculable in the weak coupling
regime,  comparison with (\ref{gs}) (for $e=0$) shows that it 
perfectly agrees
with that calculated in the strong coupling regime, implying that
there is no discontinuity in the point particle scattering amplitude
as one tunes the string coupling, for arbitrary masses of particles. 
Moreover, as expected, the gravitational interaction dominates
overwhelmingly over gauge interactions.

\section{Inclusion of charge}

To extend the results of the previous section to include gauge
interactions, the charge interaction term proportional to 
$\cosh\pi\epsilon$ has to be
retained in the kinematical factor (\ref{kappa}). This vanishes
unless $p=l$, since branes of different dimensions do not couple to
each other via gauge fields. Thus for $p=l$ :
$$ \kappa (\epsilon) \rightarrow \frac{1}{2} \left( e^{\pi \epsilon} -
4 \right)~.$$
As before, introducing $m_{p}$ and $G_4$ gives:
\begin{eqnarray}
\delta(b)~&=&~-G_4 m_p^2 \left( e^{\pi\epsilon} - 4 \right) \ln\bar  l \\
~&=&~-\left(G_4s - 4 G_4m_p^2\right)~\ln\bar  l~~.
\label{charge}
\end{eqnarray}

Let us now consider the phase shift obtained using the black
hole background, Eq. \ref{gs}. Here, inclusion of charge shows
that the coupling constant is given by $Gs -eK$. To determine
$K$, we look at the $p+1$ form potentials due to the static
brane to which the 
relativistic $p$ brane couples. In
$10$-dimensions the asymptotic value of the $p+1$ potential is given by \cite{BL}:
\begin{equation}
A_{p+1}~=~\frac{q_p}{r^{7-p}} dt\wedge
dx\wedge.....\wedge dx^{p}~,
\end{equation}
where
\begin{equation}
q_p~=~2^{5-p}~g~\pi^{\frac{5-p}{2}}\sqrt{\alpha'}^{7-p}~\Gamma\left(
\frac{7-p}{2}\right)~.
\end{equation}
The effective $U(1)$ potential due to the brane living in
$R^{10-c}\times T^c$ is obtained in two steps:
Firstly, a Kaluza Klein reduction is performed on the 10-dimensional
$p+1$ potential to obtain a one form potential in $10-p$ dimensions
due to the brane completely wrapped on $T^p$
\cite{pope}:
\begin{equation}
A_1~=~\frac{q_p}{r^{7-p}}~dt.
\end{equation}
Secondly, for the remaining $c-p$ compact directions, which are transverse
to the brane vertical reduction is performed.
For this , we stack the Kaluza-Klein reduced configurations in the
compact 
directions transverse to the brane and go to the continuum limit by
integrating over the latter, resulting in one form potentials
in $10-c$ dimensions \cite{pope}
\begin{eqnarray}
A_1~&=&~\frac{q_p\prod_{i=1}^pL_i}{V}~\int_o^\infty~\frac{d^{c-p}{r_\perp}}
{\left[r^2 + r_{\perp}^2\right]^{\frac{7-p}{2}}}dt~\\
&=&~\frac{q_p\prod_{i=1}^pL_i}{V}\int d\Omega_{c-p-1}~\int_0^\infty
\frac{r_\perp^{c-p-1}dr}{\left[r^2+
r_\perp^2\right]^{\frac{7-p}{2}}}dt
\end{eqnarray}
where $r_\perp$ refers to the transverse distance from the
$p$-brane, and $\Omega_{c-p-1}$ is the volume of the unit $(c-p-1)$
sphere, given by:
$$\Omega_{c-p-1}~=~\frac{2\pi^{\frac{c-p}{2}}}
{{\Gamma\left(\frac{c-p}{2} \right)}}~.$$
The integral is elementary, and when expressed in terms of the
$p$-brane mass and the Newton's constant given Eqs. (\ref{mass}) and
(\ref{G}) respectively, we get:
\begin{equation}
A_1~=~
\frac{4 G_{10-c}\Gamma\left(\frac{7-c}{2}\right)}{\pi^{\frac
{7-c}{2}}r^{7-c}}~m_p~dt.
\end{equation}
Comparing with the expression (\ref{amu}), we conclude that 
$$K~=~
\frac{4 G_{10-c}\Gamma\left(\frac{7-c}{2}\right)}{\pi^{\frac
{7-c}{2}}}~m_p.$$
Now, the charge of the moving brane is $e=m_p$, by the BPS
condition. 
Consequently, for $c=6$, the scattering phase shift (\ref{gs}) 
will be modified as
\begin{equation}
\delta(b)~=~-(G_4s - 4G_4 m_p^2) \ln \bar l
\end{equation}
which precisely agrees with the phase shift (\ref{charge}) obtained in the
$D$-brane side.

\section{Discussions}

We have shown that the eikonal scattering amplitude obtained by modelling
the point particles as black holes is exactly reproduced by eikonal
scattering of wrapped D-branes. In this regime only the gravitational
field at infinity is probed (as the black hole metric is linearised) and the
details of the metric are not realised. 
So, one should study the
corrections to the eikonal phase shift, as the impact parameter and velocity
are tuned to smaller values, and see whether the details begin to emerge 
\cite{DKPS1,volo}.
Our results are independent of the dimensions of the brane but
the kinematical factor $\kappa (\epsilon)$, and hence $\delta (b)$
ceases to be independent
of $p$ and $l$ once one relaxes the condition $\epsilon \rightarrow \infty$. 

We know that the type II B string theory has S-duality
group as SL(2,Z) which relates fundamental strings
to the D-strings. Under this S-duality operation,
gravitons are left invariant and NS-NS charged
fields become R-R charged fields. Our D-brane calculation
for gravitational exchange (dominant term)
matches the leading order term (eikonal limit) in the 
scattering amplitude for fundamental strings \cite{ACV}.
This confirms the S-duality symmetry. Moreover, we
have obtained the subdominant term (eikonal limit)
due to R-R charged field exchange between the two Dp-branes.
Invoking the S-duality,  we can say that this
must be also be the amplitude for NS-NS gauge field exchange
in fundamental string scattering.
In \cite{ACV}, corrections to the
eikonal fundamental string-string graviton exchange amplitude were calculated and shown to be order
$1/\bar l^2$. It will be interesting to see whether the D-brane
scattering amplitude gives the same. 

Though the fundamental string and the D-brane scattering
amplitudes are same, the relevant energy scales for eikonal
scattering are different for the two. 
For the $D$-$p$-branes to be relativistic, the condition
\begin{equation}
E \gg m_p
\label{em}
\end{equation}
must be imposed on their energy. Using the expression for the brane mass, we get
$$E \gg \frac{L^p}{g \sqrt{\alpha'}^{p+1}}~.$$
As is evident, in 10-dimensions, the mass of the D-0 brane is much larger
than the Planck mass $M_{pl}$ and the string mass
$m_s\sim1/\sqrt\alpha'$. Hence for condition (\ref{em}) to 
be realised, energies relevant for D-0 brane eikonal
scattering has to be much larger than both $M_{pl}$ and
$m_s$.
 In other words for D-0 branes to become
relativistic, we need to consider regimes where
$s g^2\alpha'\gg 1$, (In the c.m. frame 
$s=E^2$). (Note that for non-perturbative
effects of M-theory to become important, we need to have
$tg^2\alpha'\rightarrow 1$ \cite{DKPS}, and we are not probing that regime.)
On the other hand, for fundamental strings, $s\alpha'\gg 1$ gives
the eikonal limit, as considered in \cite{ACV}. 
Following \cite{shenker} we take $1/g\sqrt\alpha'$ as the energy scale in our problem.
and the condition on the compactification volume, from
Eq. \ref{em} is, 
$$ V_p \ll \sqrt \alpha'^p~,$$
which implies that the compactification radii should be sufficiently
small compared to the string scale.
Note that if we had used $1/\sqrt\alpha'$, as in
\cite{ACV} the
conditions on compactification lengths would have become $g$
dependent, and difficult to interpret.

Another interesting observation is that though the $D$-brane
scattering amplitude includes all long range closed string
exchanges, only the graviton exchange dominates in the
above kinematical regime\footnote{This was also observed in
\cite{faheem}}. This was anticipated in the black
hole calculation by 'tHooft and the weak coupling calculation
vindicates this. Our weak coupling calculations can be generalised to higher
dimensions. For example, in five non-compact dimensions, it is
easy to see
that the phase shift goes as $1/b$, which is the Green's function
for three transverse dimensions. 

One can try to examine more sophisticated compactifications e.g. on $K3$
to see whether similar conclusions hold for those
situations also \cite{DOS}. We hope to report on it in the near future.


\noindent
\begin{center}
{\bf ACKNOWLEDGEMENTS}
\end{center}
S.D. would like to thank C. Bachas, F. Hussain, R. Iengo and K. S.
Narain for useful discussions and the International Centre for
Theoretical Physics, Trieste for hospitality where part of this work was
done. A.D. would like to thank Theoretical Physics Group, Tata Institute of Fundamental
Research, Mumbai for hospitality during the course of
this work and P. Majumdar for useful discussions.

\end{document}